\documentclass[%
 aip,
 jmp,%
 amsmath,amssymb,
 reprint,%
]{revtex4-2}

\usepackage{graphicx}
\usepackage{dcolumn}
\usepackage{bm}

\begin{document}


\title{Detectability of 21-cm Power Spectra from the Epoch of Reionization using Modern Radio Telescopes}

\author{Ali Bagheri Koochaksaraei}
  \email{a.bagherikoochaksaraei@email.kntu.ac.ir}
 \affiliation{Department of Physics, K. N. Toosi University of Technology, P. O. Box 15875-4416, Tehran, Iran}
 \affiliation{PDAT Laboratory, Department of Physics, K. N. Toosi University of Technology, P.O. Box 15875-4416, Tehran, Iran}

\author{Lily Mobahat}%
 \email{leilimobahat@email.kntu.ac.ir}
\affiliation{Department of Electrical Engineering, K. N. Toosi University of Technology, P. O. Box 15875-4416, Tehran, Iran}
\affiliation{PDAT Laboratory, Department of Physics, K. N. Toosi University of Technology, P.O. Box 15875-4416, Tehran, Iran}

\author{Sara Gholamhosseinian}
 \email{sgholamhoseinian@umassd.edu}
\affiliation{Department of Physics, University of Massachusetts Dartmouth, 285 Old Westport Road, Dartmouth, MA 02747-2300, USA}
\affiliation{PDAT Laboratory, Department of Physics, K. N. Toosi University of Technology, P.O. Box 15875-4416, Tehran, Iran}

\author{Javad T. Firouzjaee}
 \email{firouzjaee@kntu.ac.ir}
\affiliation{Department of Physics, K. N. Toosi University of Technology, P. O. Box 15875-4416, Tehran, Iran}
\affiliation{PDAT Laboratory, Department of Physics, K. N. Toosi University of Technology, P.O. Box 15875-4416, Tehran, Iran}

\date{\today}

\begin{abstract}
One of the unsolved questions in astrophysics and cosmology is the development of a robust physical model to explain the thermal history of the intergalactic medium during the epoch of reionization. Various astrophysical models have been proposed, but current models remain only partially constrained by available observations. To clarify this ambiguity, hydrodynamical, semi-numerical, and N-body simulations can help solve this issue. One notable simulation from the epoch of reionization is the 21SSD simulations, which provide high-resolution results for the global 21-cm brightness temperature and the 21-cm power spectrum during the epoch of reionization. The simulation includes 45 models spanning three key astrophysical parameters of X-ray emissivity, Lyman-band emissivity, and the X-ray hard-to-soft emissivity ratio. Distinct values of these parameters result in different models of this simulation. In this work, four active radio observatories, including HERA, MWA, GMRT, and LOFAR have been considered to investigate the sensitivity and thermal noise of the 21-cm power spectrum. Each observatory has a particular sensitivity based on its instrumental characteristics, which helps assess which models lie above the forecast sensitivity thresholds of current instruments.
\end{abstract}

\maketitle


\section{Introduction}
\label{sec:sec1}
Radio waves provide vital information about astronomical objects and events, which enables us to solve cosmic mysteries beyond the reach of conventional optical telescopes. To locate radio waves originating from astronomical radio sources in the sky, one needs a radio telescope, which includes a large number of observational stations, antennae, and radio receivers \citep{panda2023unveiling}. Since radio waves are less susceptible to light pollution, radio telescopes can observe continuously throughout the day.

One of the most promising radio probes of the early gas and stars may be the hyperfine spin-flip transition of neutral hydrogen (HI) at a wavelength of approximately 21 cm. Similar to any atomic resonance line when combined with the cosmological redshift, observations of this line can be used to slice the Universe as a function of redshift (or equivalently, distance along the line of sight) \citep{barkana2016rise}.

The hydrogen 21-cm signal is expected to provide a comprehensive investigation of the early Universe. This includes the Cosmic Dark Ages (CDA), a period when the Universe was transparent, but the clouds of hydrogen collapsed slowly to form stars and galaxies, resulting in no new sources of light. The hydrogen 21-cm signal includes the Cosmic Dawn (CD), the time when the first stars and sources of light appeared, as well as the Epoch of Reionization (EoR), during which the IGM became ionized. This signal is considered to hold a lot of information about the processes that occurred in the early stages of cosmic evolution \citep{fialkov2023cosmic}. Since the signals that reach us from CD and EoR are faint and contaminated by foregrounds, thermal noise, and instrumental systematics, it is challenging to detect them due to systematic observational errors \citep{mazumder2022observing}. The birth of the first stars and galaxies caused baryonic matter, which was primarily cold neutral gas, to undergo chemical enrichment, ionization, and heating during EoR \citep{witstok2023spectroscopic}. Since the first generation of complex astrophysical objects formed during the EoR and practically all baryons in the Universe were affected by the reionization process, this era is influential in the history of our Universe \citep{lidz2015modeling}. Almost $0.2\%$ of the baryon density has been converted into stars by the end of reionization, and this fraction has continued reionizing the remaining baryonic gas. The IGM remained ionized in spite of radiative recombination because of the ionizing photons that were produced by O-type stars and active galactic nuclei (AGN).

One of the ideas is that the ionizing radiation from early black holes or galaxies is what drove reionization, even though the exact timing of the EoR is still uncertain. In this sense, the interaction of early galaxies with IGM is at the core of the reionization problem \citep{kakiichi2016high}. An important question in astrophysics, including the birth of very metal-poor stars \citep{sandford2024strong}, the features of the early galaxies \citep{munoz2022impact}, the physics of (mini-)quasars \citep{koopmans2015cosmic}, and other astrophysical study subjects, are linked to the EoR. Therefore, understanding the physics behind the EoR affects our understanding of how structures grew in the early Universe.

One of the most promising methods for studying the CDA and the EoR is observations of 21-cm emissions from HI at high redshifts \citep{zaroubi2012epoch}. Some telescopes are currently in use to measure this feeble radiation. From the high-redshift Universe, a number of current major radio observatories are LOw Frequency ARray (LOFAR) \citep{tasse2022lofar}, Murchison Widefield Array (MWA) \citep{chege2022optimising,trott2022multi}, Hydrogen Epoch of Reionization Array (HERA) \citep{abdurashidova2022first,abdurashidova2022hera}, and Giant Metrewave Radio Telescope (GMRT) \citep{swarup1991giant,pen2009gmrt,ghosh2011gmrt}. Additionally, the Square Kilometer Array (SKA) \citep{braun2019anticipated} is expected to begin its full scientific operation in the future.
 
The Universe underwent a period of darkness known as the CDA, which lasted for 400 million years and culminated in the formation of the first celestial bodies. Photon decoupling took place immediately after (or during) the cosmic recombination stage, leading to the Universe becoming transparent and the creation of the Cosmic Microwave Background (CMB) \citep{zaroubi2012epoch,wiklind2012first}.

Through gravitational instability, tiny inhomogeneities in the dark matter density field that existed during the recombination epoch (and were well-detected by the CMB) started growing, eventually generating the first stars inside galaxies. The CD arrived as these stars were formed, bringing an end to the CDA \citep{choudhury2022short}. As of right now, the GMRT has the greatest collecting area of any completely steerable telescope at all frequencies, making it the best telescope in the world in terms of sensitivity and angular resolution at low frequencies, specifically 1 GHz \citep{patra2019expanded}. A breakthrough has been made by the GMRT in many different fields, including studies of the EoR \citep{paciga2011gmrt,paciga2013simulation}, hydroxyl (OH) absorption studies of high-redshift galaxies \citep{kanekar2014spin},  giant radio sources \citep{bagchi2007giant,bagchi2014megaparsec,tamhane2015j021659,sebastian2018discovery}, HI 21-cm emission stacking studies of cosmologically distant galaxies \citep{lah2009h,kanekar2016gas,rhee2016giant,patra2019expanded}, and so forth.

LOFAR is an interferometer that has been built around Europe, especially in the northern Netherlands. LOFAR offers several unique observing capabilities and covers the mainly untapped low-frequency band from 10–240 MHz using a revolutionary phased-array architecture \citep{van2013lofar}. Throughout the phases of designing, constructing, and commissioning the project, the six Key Science Projects (KSPs) that influenced its development have essentially summarized the work done on LOFAR. These investigations include the EoR \citep{ghara2020constraining}, deep surveys of the radio sky \citep{best2023lofar}, solar and space weather \citep{carley2020radio}, cosmic magnetism \citep{beck2013lofar}, high energy cosmic rays \citep{schellart2013detecting,mulrey2020cosmic}, and the search for transients and pulsars \citep{heald2018introduction}.

MWA is a predecessor facility for SKA that is intended for advanced science using real-time imaging at frequencies between 80 and 300 MHz. The discovery of redshifted 21-cm emission from HI during the EoR \citep{paul2017study}, the detection of radio transients \citep{rajwade2017radio}, and space weather monitoring \citep{flores2021validation} are the main scientific drivers of this frequency band \citep{mitchell2010murchison}.

HERA radio telescope is currently in development to study the early Universe. HERA Phase I is intended to describe the delay power spectrum of redshifted hydrogen signal in the 100–200 MHz range \citep{pober2014next,fagnoni2021understanding}. HERA is a phased experiment designed to use the redshifted 21-cm line of HI to characterize our CD, from the full reionization of the IGM $\sim 1$ Gyr later ($z \sim 6$) to the formation of the first stars and black holes $\sim 0.1$ Gyr after the Big Bang ($z \sim 30$). HERA adds transformative capabilities to probe at other wavelengths by directly viewing the large-scale structure of the primordial IGM as it is heated and reionized, providing insights into the underlying cosmology and astrophysics of our early Universe. The special design of the HERA radio interferometer is optimized to provide high signal-to-noise measurements of redshifted 21-cm emission to identify and characterize the EoR \citep{deboer2017hydrogen}.

In addition to these observational instruments, some simulations have been employed to simulate the thermal history of the IGM, such as 21SSD \citep{semelin201721ssd}, 21-cmFAST \citep{mesinger201121cmfast}, LoReLi \citep{meriot2024loreli}, and so on. One of the main interests in astrophysics is linking simulations with observations. Some works, such as \cite{santos2008cosmic,datta2014light,cohen2020emulating} have tried to make it happen, but there is no general consensus due to the lack of accurate high-redshift observations. Therefore, the more accurate the simulations are, the narrower this gap becomes.

In Sec. \ref{sec:sec2} we discuss the EoR and its sources, such as galaxy formation and AGN, as well as its general equations. We then explain the 21SSD simulation in Sec. \ref{sec:sec3}. Detection sensitivity and the impact of thermal noise on the 21-cm power spectrum are both considered in Sec. \ref{sec:sec4}. The result is explained in Sec. \ref{sec:sec5}, where the simulated models are compared to the observational abilities of GMRT, LOFAR, MWA, and HERA radio observatories.
\section{Epoch of Reionization and its Sources}
\label{sec:sec2}
After recombination, most hydrogen in the intergalactic medium became neutral, allowing CMB photons to free-stream and initiating the Cosmic Dark Ages created shortly after the Big Bang. The ionized particles generated hydrogen and other fundamental elements, although their concentrations were modest as the temperature of the Universe decreased over time. The Universe entered the Cosmic Dark Ages before the first luminous sources gradually heated and reionized the IGM. Then the EoR began illuminating this black Universe with newly born galaxies and stars.
 
The vast majority of hydrogen atoms stayed neutral before forming stars and galaxies. However, by the time ultraviolet (UV) photons were released from the first stars in the earliest galaxies, the phase of the Universe had shifted remarkably \citep{shimabukuro2021exploring}.

Understanding the EoR is crucial because it represents the final alteration in the ionization state of the Universe, and furthermore, it had an impact on all subsequent structure development through a variety of radiative feedback effects \citep{barkana2001beginning,ciardi2005first,maio2011interplay,sobacchi2013does,wyithe2013suppressed,hutter2014visibility}.

Since the recombination of hydrogen ionized by their young stellar populations produce the intrinsically brightest emission line, $Ly\alpha$, these galaxies should be great sources of this emission \citep{dijkstra2014lyalpha,heintz2023extreme,witten2024deciphering}.

One of the main objectives of high-redshift extragalactic astrophysics is to identify and classify the sources of the EoR. Currently, several sources have been suggested to be involved in reionization consisting of dwarf galaxies, massive galaxies, AGN, globular clusters (GCs), stellar-mass black holes, accretion shocks, and dark matter annihilation and decay \citep{katz2019tracing}. The most likely candidates to be the luminous sources that reionized our Universe are AGN and star-forming galaxies \citep{hassan2019identifying}.

Evidence points to actively star-forming galaxies being the source of the majority of the necessary ionizing photons during this period \citep{robertson2013new,puchwein2019consistent,meyer2019role}, although there is currently some disagreement regarding the potential effects of AGN on the reionization at $z \sim 6$ \citep{giallongo2015faint,chardin2016large,d2017contribution,meyer2019role}.

There is a further essential aspect of the photons from stars that contribute to reionization. It is necessary to remember that not every photon created within a galaxy will be able to escape into the IGM that surrounds it.  A fraction will be absorbed by dense clouds of HI in the interstellar medium (ISM) of the galaxy. At high redshifts, there are almost no observational limitations, making the challenge of modeling this impact extremely difficult. Therefore, the mean fraction of photons that escape into the IGM from halos of mass M generated at $t_{form}$, is typically used for determining this uncertainty \citep{choudhury2022short}. Since faint star-forming galaxies are discovered to be common and may have large ionizing photon escape fractions, they are highly desirable candidates as being the main contributors to cosmic reionization \citep{naidu2020rapid}.

Compared to typical reionizing galaxies, AGN are more strongly biased toward massive halos, which leads to higher clustering and later formation timeframes \citep{hassan2018constraining}. The model used in \cite{finkelstein2019conditions} favors a small AGN contribution toward the end of reionization, with AGN accounting for around one-third of the total budget of ionizing photons at z = 6. Even though this contribution is higher than many prior estimates, it is still consistent with the few existing observations of He II reionization, for which there are substantial uncertainties.

If ionizations by secondary electrons are taken into consideration, quasars may also contribute significantly to reionization, particularly at $z\gtrsim 8$. At $z\lesssim 6$, galaxies take over as the primary generators of reionization \citep{dayal2018early}.

Dark matter can annihilate or decay into standard model particles in a number of scenarios. These particles can then provide energy to the IGM by ionization, heating, or other methods. When structure creation begins and dark matter collapses into dense halos, the annihilation rate, which grows as the square of the density, rises significantly, potentially producing a huge amount of energy injection in the EoR \citep{liu2016contributions}.

Brightness temperature, $T_{b}$, is a measure of the observed intensity of radiation from an astronomical object, expressed as the temperature of a black body emitting the same specific intensity at a given frequency \citep{rybicki1991radiative}. The 21-cm global brightness temperature can be used to measure the 21-cm power spectrum or as a sky-averaged global signal. It is possible to decompose the spatial fluctuations in brightness temperature within a 3D comoving volume into various Fourier modes, as provided by 
\begin{equation}
    \delta T_b(\overline{x})=\int \frac{d^3k}{(2\pi)^3}\delta T_b(\overline{k})e^{i\overline{k}\cdot\overline{x}},
    \label{eq:eq3}
\end{equation}
where $\delta{T_b}(\overline{k})$ is the fourier conjugate of $\delta{T_b}(\overline{x})$. The expectation value of this quantity, $\delta{T_b}(\overline{k})$, can be expressed as follows \citep{choudhury2022extracting}
\begin{equation}
    \left\langle \delta{T_b}(\overline{\mathbf{k}}) \delta{T_b}^{\ast}(\overline{\mathbf{k}^{\prime}}) \right\rangle = (2\pi)^{3} \delta_{D}(\mathbf{k} - \mathbf{k}^{\prime})P_{21}(\overline{\mathbf{k}})
    \label{eq:eq4}
\end{equation}
Here, $P_{21}(\mathbf{k})$ is the 21-cm 3D power spectrum and $\delta$ is the Dirac delta function. Note that although the wave vector can have both positive and negative values, $\delta{T_b}^*(\mathbf{k}) = \delta{T_b}(-\mathbf{k})$, since $\delta{T_b}(\mathbf{x})$ is a real field. It is also common to define the dimensionless power spectrum as $\Delta_{21}^{2}(k) = \mathbf{k}^{3}P_{21}(\mathbf{k})/(2\pi)^{2}$ where $k$ is the wavenumber.

Moreover, according to the formal solution of the radiative transfer equation, the following formula can be obtained for the brightness temperature \citep{Kuhlen_2006}
\begin{equation}
    \delta T_{b}(z,\nu) = \frac{T_{S}-T_{\gamma}}{1+z}\left(1-e^{-\tau(z,\nu)}\right) \approx \frac{T_{S}-T_{\gamma}}{1+z} \tau,
    \label{eq:eq1}
\end{equation}
where $T_{\gamma}$ defines the background temperature of radio photons. Since the CMB acts as the source of background radiation, $T_{\gamma} = 2.73(1+z)K$ is the CMB temperature. Moreover, $T_{S}$ is the spin temperature describing the relative occupancy of the two spin states, and $\tau$ is the optical depth of the IGM at frequency $\nu$ and redshift $z$, expressed as \citep{Madau_1997}
\begin{align}
    \begin{aligned}
        \tau(z,\nu) & =\frac{3}{32 \pi} \frac{h c^3 A_{10}}{k_B T_S \nu^2} \frac{x_{\mathrm{HI}} n_H}{(1+z)\left(\mathrm{d} v_{\|} / \mathrm{d} r_{\|}\right)} \\
        &\approx 9.2\times10^{-3}(1+\delta)(1+z)^{3 / 2} \\
        &\times\frac{x_{\mathrm{HI}}}{T_S}\left[\frac{H(z) /(1+z)}{\mathrm{d} v_{\|} / \mathrm{d} r_{\|}}\right],
    \end{aligned}
\label{eq:eq2}
\end{align}
where $h$ is the Planck constant, $A_{10}$ is the spontaneous emission coefficient of the 21-cm transition, $x_{\mathrm{HI}}$ is the fraction of HI, $n_{H}$ is the hydrogen number density, $\delta$ is the factor accounting for the overdensity of baryons, $H(z)$ represents the evolution of the Hubble constant, and $\mathrm{d} v_{\|} / \mathrm{d} r_{\|}$ is the velocity gradient along the line of sight. According to Eq. \eqref{eq:eq2}, the spin temperature determines whether the 21-cm signal appears in emission or absorption.

\begin{table*}
    \centering
    \caption{\centering Parameters describing arrays}
    \label{tab:tab1}
    \renewcommand{\arraystretch}{1.5}
    \hspace{-1.3cm}
    \begin{tabular}{|c|c|c|c|c|}
        \hline
         & \textbf{HERA} & \textbf{GMRT} & \textbf{LOFAR} & \textbf{MWA} \\
        \hline
        $\mathbf{R_{max} \; [km]}$ & $0.876 \citep{Ewall_Wice_2016}$ & $12.5$ \citep{girish2023progression} & $2$ \citep{gehlot2019first} & $1.5$ \citep{girish2023progression} \\
        \hline
        $\mathbf{A_{e} \; \left[10^{-3} km^{2}\right]}$ & $0.551$ \citep{deboer2017hydrogen} & $1.6$ \citep{bagla2010h} & $0.00533$ \citep{gehlot2024transient} & $0.02$ \citep{beardsley2015adding} \\
        \hline
        $\mathbf{B \; [MHz]}$ & $100$ \citep{perini2022radio} & $32$ \citep{mangla2022study} & $96$ \citep{gehlot2019first} & $6$ \citep{beardsley2015adding} \\
        \hline
        $\mathbf{\Omega(sr) \; }$ & $0.0076$ \citep{sinclair2015study} & $0.011$ \citep{mazumder2020characterizing} & $0.069$ \citep{braun2013understanding} & $0.274$ \citep{tremblay2017first} \\
        \hline
        $\mathbf{N \; }$ & $350$ \citep{raste2021implications} & $30$ \citep{paciga2011gmrt} & $576$ \citep{gehlot2024transient} & $112$ \citep{beardsley2015adding} \\
        \hline
        $\mathbf{T_{rec} \; [K] }$ & $160$ \citep{fagnoni2021design} & $295$ \citep{gupta2014gmrt} & $140$ \citep{hassan2017epoch} & $200$ \citep{bowman2007field} \\
        \hline
    \end{tabular}
\end{table*}

The power spectrum in the standard model is naturally and necessarily not exactly what exists and is observed in nature, for various reasons, as discussed. Therefore, we need simulations that can account for and reduce this difference or discrepancy, allowing us to achieve a better model of the IGM history and, more generally, the thermal history of the universe, or to place better constraints on them.

In the next section, we explain which simulation we selected and the reasons behind our choice.

\section{the 21SSD Simulation}
\label{sec:sec3}
One of the main and important simulations of the EoR is the 21SSD simulation \citep{semelin201721ssd}, which the 21SSD simulation provides 21-cm brightness-temperature lightcones for 45 models, covering a selected 3D parameter space. This hydrodynamical simulation uses $1024^3$ particles, with half representing gas and half representing dark matter, in a 200 Mpc.$h^{-1}$ cube. In addition, a $512^{3}$ fixed grid has been used to evaluate the local $Ly\alpha$ flux and the 21-cm brightness temperature. The radiative-transfer calculation follows $4 \times 10^8$ photon packets between consecutive snapshots during the time interval between two snapshots of the simulation, which corresponds to $10^{7}$ years. The 21SSD simulation includes different values for three astrophysical parameters. These parameters are X-ray emissivity, $f_{x}$, the Lyman band emissivity, $f_{\alpha}$, and the X-ray hard-to-soft emissivity ratio, $r_{H/S}$.

One of the main reasons for the heating of the IGM is its X-ray emissivity. Explaining X-ray emission at high redshifts is quite challenging due to the uncertainties surrounding the nature of objects at these redshifts. However, X-ray measurements provide a new approach to determining the SFR of a galaxy \citep{Grimm_2003}. The relationship between SFR and X-ray luminosity at high redshifts can be expressed using the parameter $f_x$ \citep{list2020unified}

\begin{equation}
    L_{x} = 3.4 \times 10^{40} f_{x}\left(\frac{SFR}{1\,M_{\odot}\,\mathrm{yr}^{-1}}\right) \; erg \ s^{-1}.
    \label{eq:eq5}
\end{equation}

Another parameter is the Lyman band emissivity. The energy emitted by a star population as being between frequencies $\nu_{1}$ and $\nu_{2}$ can be defined as
\begin{equation}
    E(\nu_{1}, \nu_{2}) = \int_{\nu_{1}}^{\nu_{2}}\int_{M} \xi(M) L(M, \nu) T_{\text{Life}}(M) dM d\nu,
    \label{eq:eq6}
\end{equation}
where $M$ is the stellar mass, $\xi(M)$ is the Initial Mass Function (IMF) of the stellar population, $T_{\text{Life}}(M)$ is the lifetime of the star, and $L(M, \nu)$ is the energy emitted by the star at frequency $\nu$ per time. The proportionality between the energy produced in the simulation, $E_{eff}$, and the theoretical energy emission in the pertinent frequency range, E, is then defined below
\begin{equation}
    f_{\alpha} = \frac{E_{eff}(\nu_{\alpha}, \nu_{\text{limit}})}{E(\nu_{\alpha}, \nu_{\text{limit}})},
    \label{eq:eq7}
\end{equation}
where $f_{\alpha}$ is the Lyman band emissivity, $\nu_{\alpha}$ is the $Ly\alpha$ frequency, and $\nu_{\text{limit}}$ is the limit frequency of the Lyman band.

X-ray binaries and AGN produce substantial X-ray emission, while they differ in their spectral characteristics and effectiveness in heating the IGM \citep{fialkov2014observable}. The hard X-rays have longer mean free paths and are therefore less efficient at local IGM heating. However, the X-ray photons produced by AGN are soft X-rays have shorter mean free paths and deposit their energy more locally \citep{fragos2013x}. The X-ray hard-to-soft ratio, which defines the X-ray production efficiency in simulations is the ratio of X-rays produced from binaries to X-rays produced from all sources including binaries and AGN
\begin{equation}
    r_{H/S} = \frac{f_{X}^{(XRB)}}{f_{X}^{(XRB)} + f_{X}^{(AGN)}},
    \label{eq:eq8}
\end{equation}
where $f_{X}^{(XRB)}$ is the X-ray emissivity of X-ray binaries and $f_{X}^{(AGN)}$ is the X-ray emissivity of AGN.

In order to compare different models of this simulation to reality, the utilization of radio observatories becomes crucial. Estimating the sensitivity and thermal noise associated with each observatory marks the initial phase of this process. Furthermore, the subsequent section not only offers an overview of the estimation methods but also delineates the instrumental characteristics of each observatory.

\section{Detection sensitivity and Thermal noise}
\label{sec:sec4}
The sensitivity of a radio telescope is a measure of its ability to detect weak signals from astronomical sources. The sensitivity with $1\sigma$ error in the 21-cm power spectrum is expressed by the following formula \citep{cole2021small}
\begin{align}
    \begin{aligned}
        \sqrt{\frac{k^{3}\delta P_{21}}{2\pi^{2}}} &\sim \frac{0.1\,\mathrm{mK}}{\epsilon^{1/4}f_{cov}}\left(\frac{k}{0.04\,\mathrm{Mpc}^{-1}}\right)^{3/4} \left(\frac{T_{sky}}{10^{4}\,\mathrm{K}} \frac{2\,\mathrm{km}}{R_{max}}\right) \\
        &\times \left(\frac{10\,\mathrm{MHz}}{B}\right)^{1/4} \left(\frac{1000\,\mathrm{hr}}{t_{int}}\right)^{1/2} \left(\frac{1+z}{50}\right),
    \end{aligned}
    \label{eq:eq9}
\end{align}
where $\epsilon$ is the frequency in $k$ that the data is binned, $T_{sky} \equiv 180(1420/(180(1+z)))^{-2.6}$ K is the galactic synchrotron foreground temperature at the frequency of observation \citep{yang2022impact}, $R_{max}$ is the array radius, $f_{cov} \equiv N A_{e}/(\pi R_{max}^{2})$ is the array covering factor ($A_{e}$ is its effective area) \citep{furlanetto2006effects}, $B$ is the bandwidth of the observation, and $t_{int}$ is the total integration time.

In this paper, we have considered HERA, GMRT, LOFAR, and MWA arrays to investigate their ability to detect the power spectrum of 21SSD simulations. Each of these arrays operates in a certain range of redshift and wavenumber. In order to explore the early Universe, including the birth of the first stars and black holes following the Big Bang until the entire reionization of the IGM around 1 billion years later, HERA is designed to detect or statistically constrain the redshifted 21-cm signal of H\text{I}. In South Africa, it is comprised of a 350-element interferometer with 14-meter parabolic dishes that observe between 50 and 250 MHz \citep{deboer2017hydrogen}. It operates in the wavenumber between $0.01$ and $10$ and the redshift range of $6$ to $13$.

The GMRT consists of 30 parabolic dishes, each 45 meters in diameter, and operates over a wavenumber range of $0.1$ to $1$ and a redshift range of $1.5$ to $8.5$ \citep{Ali_2008}.. LOFAR is an interferometric array extending across Europe, operating in the 10–240 MHz frequency range. The core receiving components of LOFAR consist of two types of small, cost-effective antennae that collectively span the $30$ to $240$ MHz operating range. These antennae are systematically arranged into 48 distinct stations. It is one of the first radio observatories with automated processing pipelines to provide its user community with fully calibrated science products \citep{van2013lofar}. It works in the wavenumber range of $0.01$ to $10$ and the redshift range of $6$ to $19$.

The MWA is a radio telescope is especially designed to operate between the frequency range of 80 and 300 MHz. The MWA features dual-polarisation dipole antennas, which have a nominal collecting area of about one square meter at a frequency of 150 MHz. Sixteen antennas are systematically arrayed in a 4×4 grid, a uniform layout repeated throughout the aperture array. The wide field of view of the MWA is a financially feasible method of increasing sensitivity to  EoR power spectrum fluctuations through increasing the area of the sampled sky. The range of the MWA lies between wavenumbers of 0.001 up to 10, and redshifts between 3.5 up to 19 \citep{tingay2013murchison}.

\cite{parsons2012sensitivity}, which considered the parameters of the Precision Array for Probing the Epoch of Reionization (PAPER) radio observatory, developed the following approach to assess the effect of noise on the power spectrum and illustrate the impact of thermal noise on the 21-cm power spectrum.
 
The equation below defines the squared visibility, extending the definition of visibility within the flat-sky approximation to include a similar Fourier transform along the frequency axis 
\begin{align}
    \tilde{V}^2_{21}(u, v, \eta) = \left(\frac{2k_B}{\lambda^{2}}\right)^2
    {\frac{\Omega B}{X^2Y}}{\frac{2\pi^2}{k^3}}{\Delta^{2}_{21}(k)};
    \label{eq:eq10}
\end{align}
where $\Omega$ represents the solid angle of the primary beam of one element, measured in steradians. The subscript "21" indicates that these values are calculated for the reionization 21-cm emission. $X^{2}Y$ is a scalar used to convert observed units to cosmological distances in $h^{-1} Mpc$ ($X$ translates from angle on the sky to transverse distances and $Y$ converts from bandwidth to line-of-sight distance) \citep{pober2013baryon}. $X^2Y$ is defined as below
\begin{align}
    X^2Y \approx 540\left(\frac{1+z}{10}\right)^{0.9} \; \frac{h^{-3}\mathrm{Mpc}^3}{\mathrm{sr}\,\mathrm{Hz}}.
    \label{eq:eq11}
\end{align}

A white-noise signal having an rms brightness temperature of $T_{N,rms}$ is produced by thermal fluctuations. In practical terms, this signal roughly corresponds to the sky temperature for 21-cm instruments. The thermal noise adds a component to the rms amplitude of the visibility $\tilde{V}_N$ that is defined as 
\begin{align}
    \tilde{V}_N = \frac{2k_B}{\lambda^{2}}
    {T_{N,rms}} 
    {\Omega} 
    {B}.
    \label{eq:eq12}
\end{align}
The following equation gives the power spectrum of the thermal noise obtained from one integration of a dual-polarization baseline. Noise contribution to the dimensionless power spectrum, $\Delta^{2}_{N}(k)$, is shown by the following equation \citep{pober2013baryon}.
\begin{align}
    \Delta^{2}_{N}(k) \approx X^{2}Y \frac{k^{3}}{2\pi^{2}} \;
    \frac{\Omega}{2t_{int}} \;
    T^{2}_{sys},
    \label{eq:eq13}
\end{align}
This equation can be derived by considering ${T^2}_{N,rms} = {T^2}_{sys}/Bt$, Eq. \eqref{eq:eq10} and Eq. \eqref{eq:eq12}. The outcome of array noise, including the effects of sampling redundancy, is \citep{parsons2012sensitivity} 
\begin{align}
    \begin{aligned}
        \Delta^{2}_{N}(k) & \approx 58\left[\frac{k}{0.1\,h\,\mathrm{Mpc}^{-1}}\right]^{\frac{5}{2}}\times\left[\frac{6\,\mathrm{MHz}}{B}\right]^{\frac{1}{2}}\times\left[\frac{1}{\Delta\ln{k}}\right]^{\frac{1}{2}} \\
        &\times\left[\frac{\Omega}{0.76\,\mathrm{str}}\right]\times\left[\frac{T_{sys}}{500\,\mathrm{K}}\right]^{2}\times\left[\frac{6\,\mathrm{hrs}}{t_{per-day}}\right]^\frac{1}{2} \\
        &\times\left[\frac{120\,\mathrm{days}}{t}\right]\times\left[\frac{32}{N}\right]\times\left[\frac{10^4f_0}{f}\right]^\frac{1}{2}{\mathrm{mK}^2};
    \end{aligned}
    \label{eq:eq14}
\end{align}
where $t_{per-day}$ stands for the amount of time a baseline samples a $(u, v, \eta)$-mode per day and N is the number of antennae, and also data can be divided into equally spaced intervals of $\Delta\ln{k}$. A factor of $[f/f_0]^{1/2}$ has been used to incorporate the effects of sampling redundancy in the noise power spectrum. For a filled circular aperture, the instantaneous redundancy of an array is $f = NA_e/2 = Nf_0$. Furthermore, the system temperature is defined as $T_\mathrm{sys} = T_\mathrm{sky} + T_\mathrm{rec}$, where $T_{rec}$ is the receiver temperature and the data for each radio observatory can be found in Table \ref{tab:tab1}. The sky temperature has been assumed to be \citep{thompson2017interferometry,raste2021implications}
\begin{align}
    T_{sky}(z) = 60\,\mathrm{K} \left(\frac{300\,\mathrm{MHz}}{\nu(z)}\right)^2,
    \label{eq:eq15}
\end{align}
where $\nu(z) = 1420/(1+z)$.

The parameters required to calculate the sensitivity and thermal noise, pertaining to the characteristics of the HERA, GMRT, LOFAR, and MWA radio observatories, are provided in Table \ref{tab:tab1}.

In Sec. \ref{sec:sec5}, according to Table \ref{tab:tab1}, all the 45 21SSD simulated spectra are compared with forecast sensitivities and published upper limits for each of them.

\section{Results}
\label{sec:sec5}
Among the current four main radio observatories, LOFAR, GMRT, MWA, and HERA, which try to probe high-redshift Universe, HERA is the most promising due to the great sensitivity and capacity to distinguish thermal noise from the foreground signal. The investigation of the abilities of each radio observatory is predominant because they can potentially deepen our knowledge of the evolution of the Universe through cosmic time.

It needs to be determined which astrophysical model prevailed during this era, which is dependent on different coefficients of X-ray emissivity, Lyman band emissivity, and X-ray hard-to-soft ratio based on the 21SSD simulation. Finding the correct combination of these coefficients depends on the ability of radio telescopes to put constraints on the 21-cm power spectrum.

Different simulations can be utilized to explore these astrophysical models, but this work is mainly devoted to the results of the 21SSD hydrodynamical simulation. Considering today’s technology and radio telescope limitations, we evaluate the 45 models represented by 21SSD to identify those that are most consistent with the thermal history of the universe or at least put an upper-bound for the 21-cm global signal 21SSD simulations. We aim to support this by examining the power spectrum of each model and incorporating the observational limits from these radio telescopes.

First of all, numerical simulations that are employed incorporate some required approximations that can affect the accuracy of the simulated matter power spectrum. These methods establish the equations of motion of the dark matter particles together with baryonic matter in a discrete format. For example, rather than taking into consideration a continuous distribution of dark matter particles, the simulation may depict this distribution as some significant but discrete number of particles \citep{Calder_2002}.

Inadequate particle resolution will lead to under-sampling of the matter distribution along with the loss of small-scale structures, resulting in a biased power spectrum of matter \citep{springel2005cosmological,vogelsberger2014properties}. The box size of the simulation is another effective factor, in addition to simulation grid resolution, that will lead to the accuracy of the power spectrum that has been simulated to be changed. Lower resolution or a smaller box may not be able to capture all scales that are in the power spectrum of $\Lambda$CDM, resulting in discrepancies \citep{heitmann2013coyote, heitmann2019outer}.

Furthermore, baryonic influences such as star formation, supernovae and black hole feedback, as well as cooling, can change the distribution of matter and, as such, the matter power spectrum. The use of baryonic influences within the simulation can produce different outcomes compared to simulations that use dark matter alone \citep{chisari2018impact}. Another aspect that plays a crucial role is the simulation's initial conditions such as the amplitude as well as the distribution of density fluctuation that can substantially impact the power spectrum that is obtained. The power spectrum of evolved matter can be quite different with minimal variations in the simulation's initial conditions. Some of such variations can be the change of amplitude in the density fluctuations or the development of some additional features within the initial density field such that the matter clustering on different scales can be varied. This happens as a departure from the $\Lambda$CDM matter power spectrum, mostly prominent over the smaller scales wherein non-linear phenomena set in \citep{audren2011non,valkenburg2015initial}.

In addition to the global signal bound, we also compare our results with upper limits on the 21-cm power spectrum from HERA, GMRT, LOFAR, and MWA. In all cases, an upper limit represents the maximum fluctuation power consistent with the data when no detection is claimed, under the assumption that the true power is non-negative. Based on \cite{Abdurashidova_2023}, the current limits of HERA come from a conservative, noise-dominated analysis that removes foreground-contaminated modes and systematics, applies a non-negative prior, and uses extensive simulations and statistical tests to verify that any signal loss is negligible. GMRT removes smooth-spectrum foregrounds via singular value decomposition, corrects for signal loss using injected mock signals, and uses cross-night spectra to suppress noise \citep{paciga2013simulation}. LOFAR improves gain calibration, mitigates foregrounds with Gaussian Process Regression, and applies optimally weighted power-spectrum estimation, achieving a bound about eight times tighter than previous LOFAR results \citep{Mertens_2020}. MWA Phase II applies refined calibration, strict data-quality control, and improved analysis methods to deep epoch-of-reionization observations, reporting limits at multiple redshifts with improved sensitivity over earlier MWA work \citep{Li_2019}. Together, these upper limits provide a robust set of observational constraints for testing and refining astrophysical models.

This work presents a comparison between simulated outcomes and observational upper limits at specific redshifts that correspond to those given in the respective references to each radio telescope. We are focusing in particular on redshift $z=7.1$ for MWA, $z=8$ for HERA, $z=8.6$ for GMRT, and $z=9.1$ for LOFAR. This redshift-specific strategy was adopted to enhance the reliability as well as direct relevance of our comparisons. This methodology can be generalized to encompass the entire range of redshifts over which each instrument can be used.

\begin{figure}
    \centering
    \includegraphics[width=\linewidth]{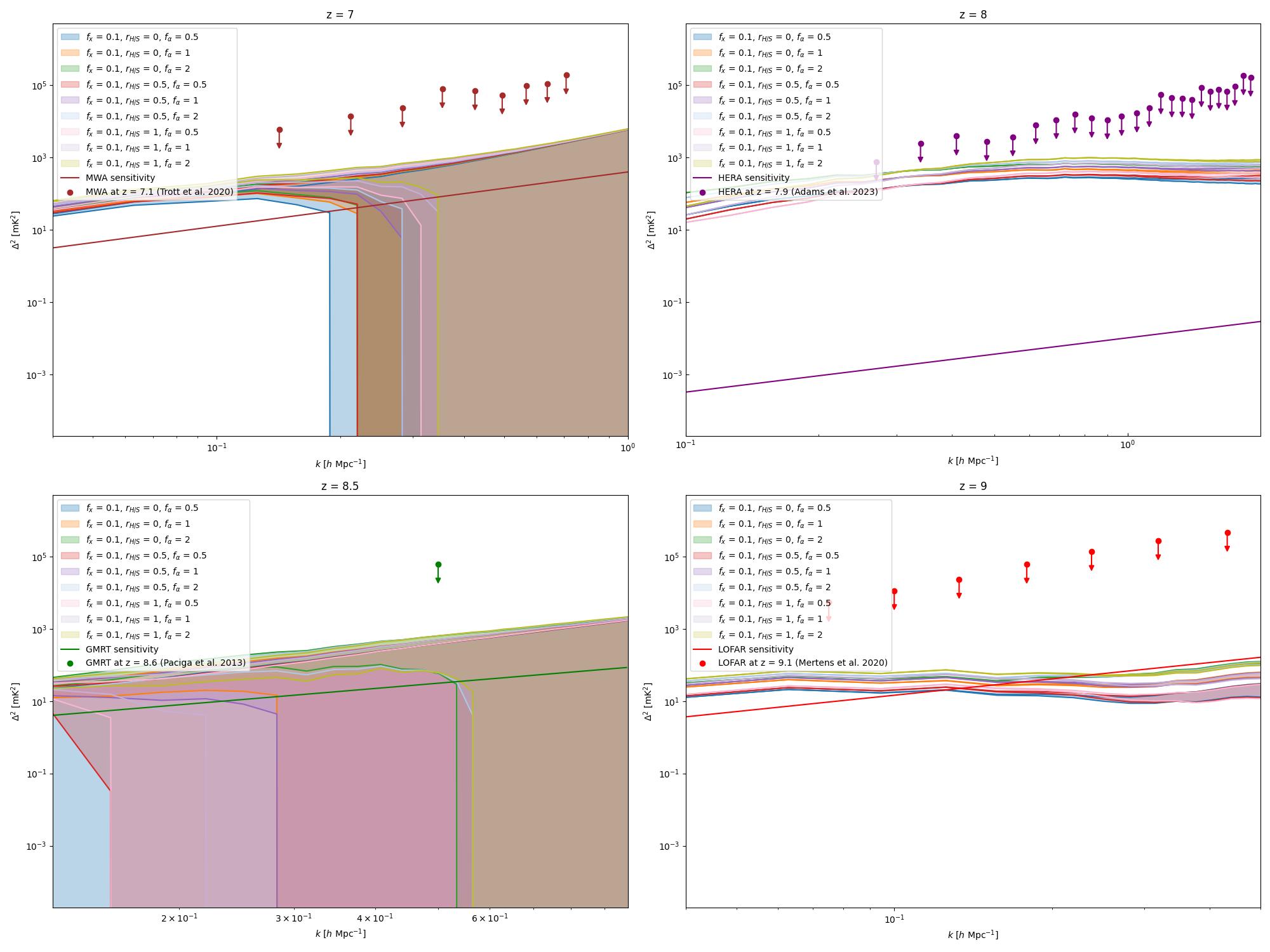}
    \caption{Comparison of 21SSD model amplitudes with forecast sensitivity curves and published upper limits at $f_{x} = 0.1$ with the sensitivity thresholds and upper limits of the four radio telescopes.}
    \label{fig:fig1a}
\end{figure}
\begin{figure}
    \centering
    \includegraphics[width=\linewidth]{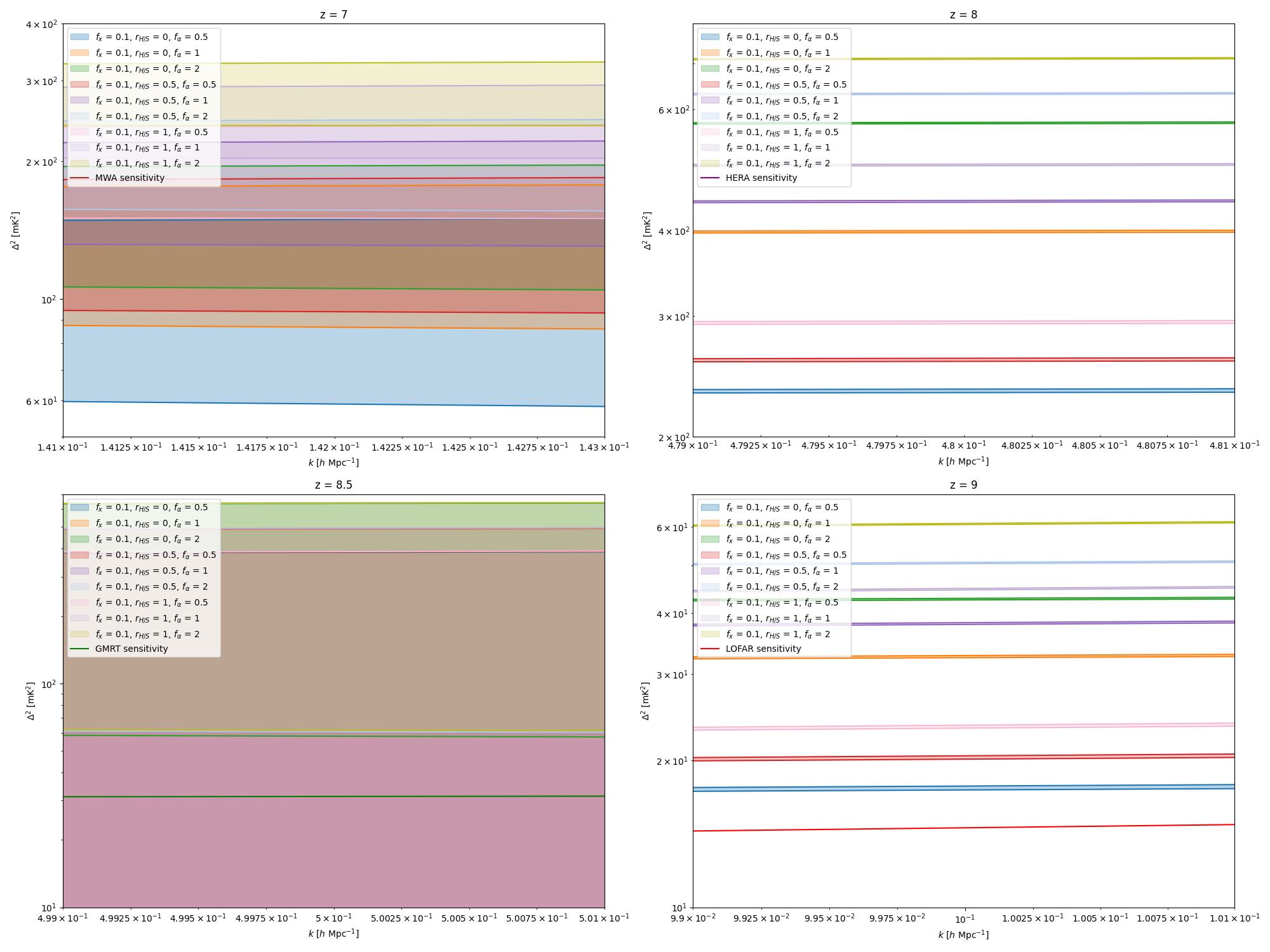}
    \caption{Thermal-noise bands indicate the expected uncertainty at the representative wavenumber of the 21SSD models at $f_x = 0.1$.}
    \label{fig:fig1b}
\end{figure}
Fig. \ref{fig:fig1a} illustrates the detectability of the 21SSD model at a fixed $f_x = 0.1$. Meanwhile, Fig. \ref{fig:fig1b} evaluates the ability to distinguish between different models once they have been detected in the presence of thermal noise. Each of the four panels in both figures corresponds to a specific radio telescope at the redshifts provided earlier. From Fig. \ref{fig:fig1a}, it is evident that the simulated spectra lie above the forecast sensitivity thresholds since they are above the sensitivity threshold of each telescope. This comparison should be interpreted as a consistency or constraint test on the 21-cm power spectrum for each radio telescope. However, it is important to note that LOFAR is sensitive to these models only within a restricted wavenumber interval; it demonstrates the most extensive detection coverage for a model with $ r_{H/S} = 1 $ and $f_\alpha = 2 $ up to $k = 0.142 \ h\,\mathrm{Mpc}^{-1}$, while its coverage is least effective for the model with $r_{H/S} = 0$ and $f_\alpha = 0.5$ up to $k = 0.1 \ h\,\mathrm{Mpc}^{-1}$.

\begin{figure}
    \centering
    \includegraphics[width=\linewidth]{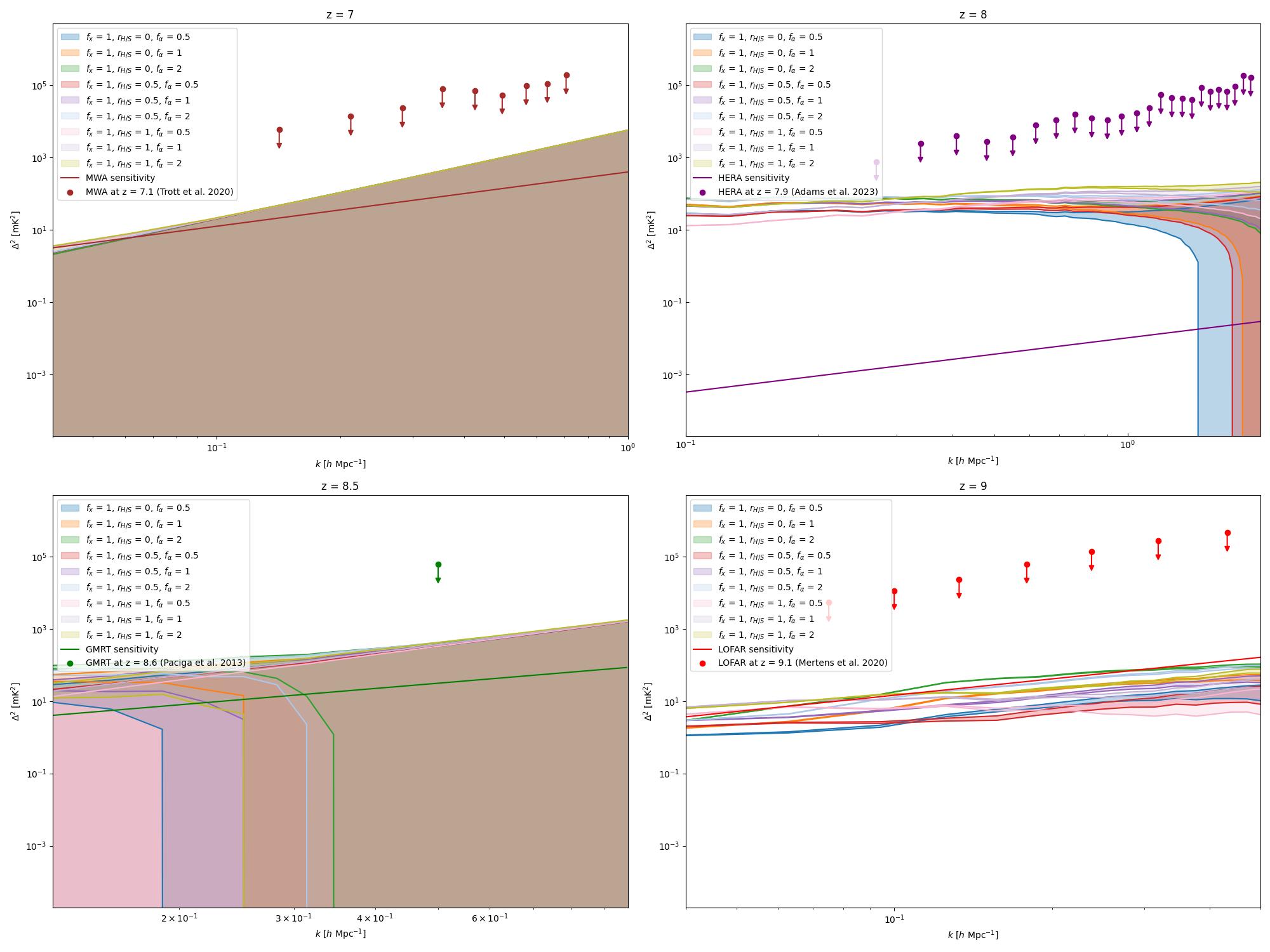}
    \caption{Comparison of 21SSD model amplitudes with forecast sensitivity curves and published upper limits at $f_{x} = 1$ with the sensitivity thresholds and upper limits of the four radio telescopes.}
    \label{fig:fig2a}
\end{figure}
Similar to Fig. \ref{fig:fig1a}, Fig. \ref{fig:fig2a} illustrates the detectability of the 21SSD models at $f_x = 1 $, considering the observational upper limits on the power spectrum of each radio telescope at their corresponding redshift and wavenumber. It shows that the MWA is sensitive over a limited part of the wavenumber range, as they lie above its sensitivity line over a large area within its wavenumber range. However, models with $r_{H/S} = 0$ and $f_\alpha = 2$, as well as those with $r_{H/S} = 0$ and $f_\alpha = 0.5$, lie below the sensitivity line before $k = 0.5 \ h\,\mathrm{Mpc}^{-1}$. HERA and GMRT demonstrate strong detectability for all models since their sensitivity lines are below the models. In contrast, LOFAR does not perform as well in detecting the 21SSD models at this redshift. It can only detect the model with $r_{H/S} = 0$ and $f_\alpha = 2$ up to $k = 0.3 \ h\,\mathrm{Mpc}^{-1}$, the model with $r_{H/S} = 0.5$ and $f_\alpha = 1$, and the model with $r_{H/S} = 1$ and $f_\alpha = 2$ up to $k = 0.1 \ h\,\mathrm{Mpc}^{-1}$. All other models are nearly undetectable, as they are entirely or significantly below LOFAR's sensitivity.

\begin{figure}
    \centering
    \includegraphics[width=\linewidth]{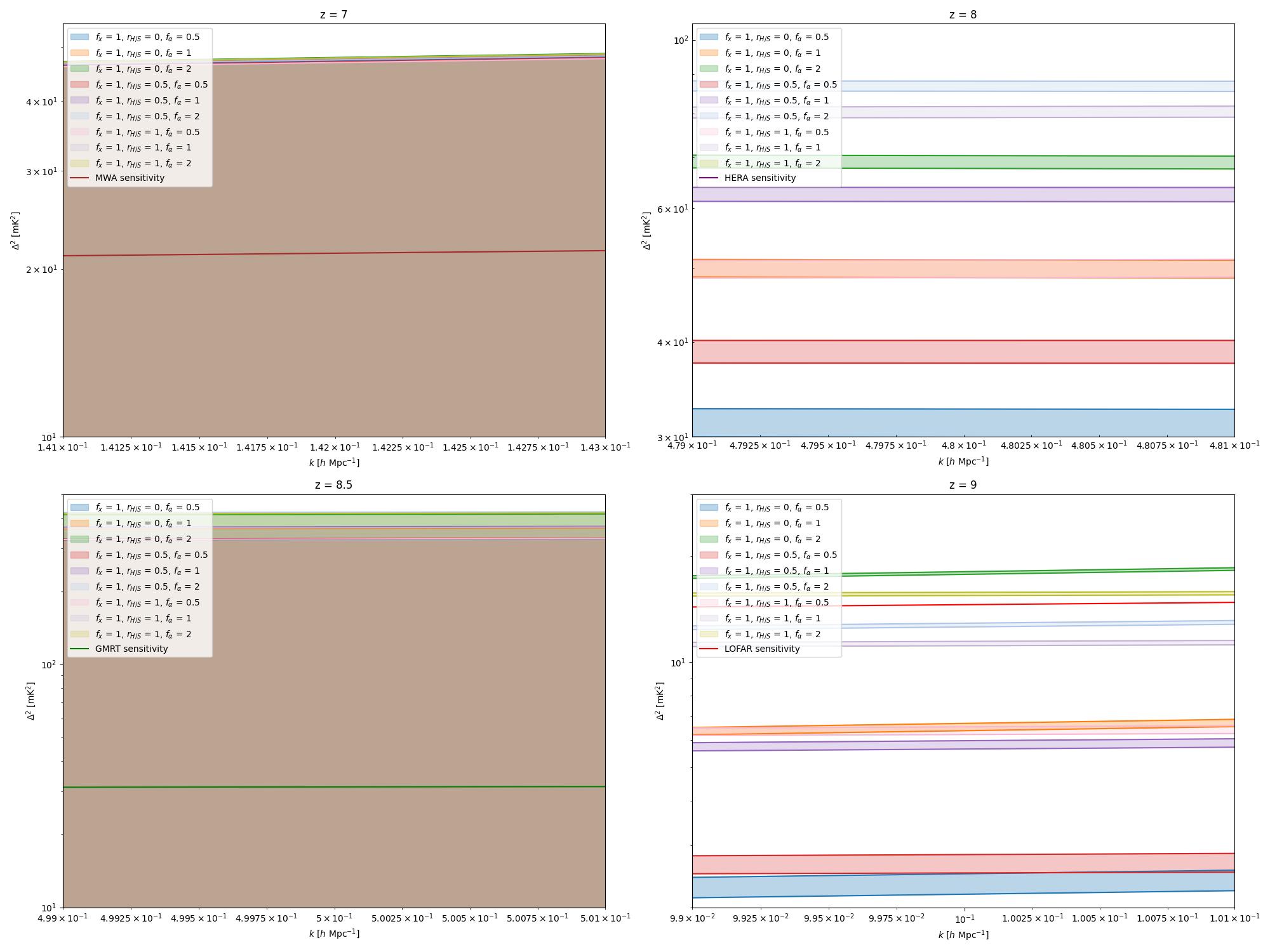}
    \caption{Thermal-noise bands indicate the expected uncertainty at the representative wavenumber of the 21SSD models at $f_x = 1$.}
    \label{fig:fig2b}
\end{figure}
Fig. \ref{fig:fig2b} shows how thermal noise from each radio telescope can impact their ability to distinguish different signals at $f_x = 1$. Similar to the situation at $f_x = 0.1$, both the MWA and GMRT struggle to distinguish any of the models from one another due to their high thermal noise levels at $k = 0.142 \ h\,\mathrm{Mpc}^{-1}$ and $k=0.5 \ h\,\mathrm{Mpc}^{-1}$, respectively. In contrast, HERA demonstrates an impressive capability to differentiate all the models, as indicated in the upper right panel, which is very promising given that it can also detect signals at $k = 0.48 \ h\,\mathrm{Mpc}^{-1}$. Although LOFAR is not able to detect all the signals across the entire range of wavenumbers, it shows visual separation among the model predictions at a specific wavenumber of $k = 0.1 \ h\,\mathrm{Mpc}^{-1}$. More specifically, within this range, its thermal noise prevents it from distinguishing between just two candidates with $r_{H/S} = 0$ and $r_{H/S} = 0.5$, both having the same $f_\alpha$ value of $0.5$.

\begin{figure}
    \centering
    \includegraphics[width=\linewidth]{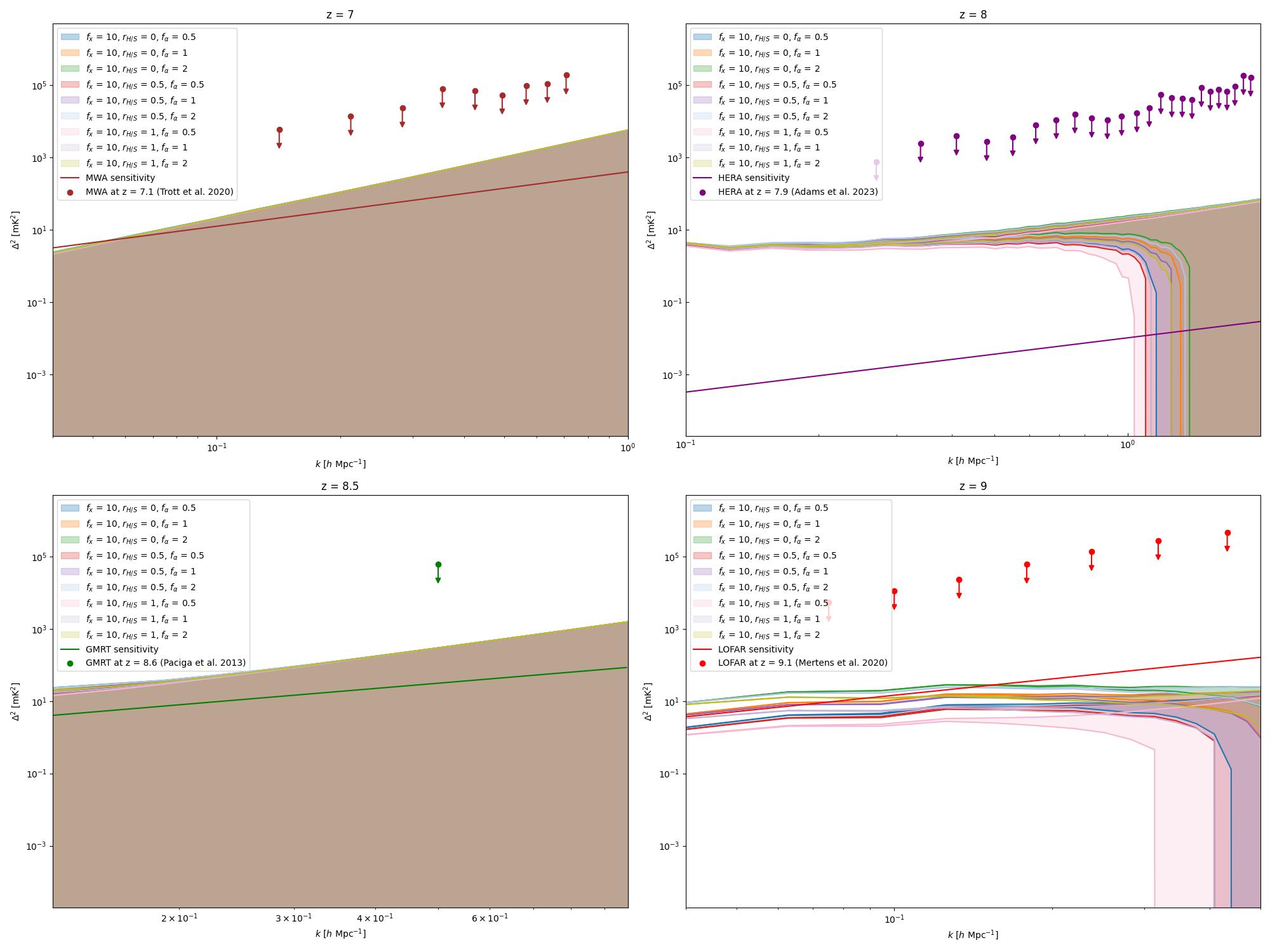}
    \caption{Comparison of 21SSD model amplitudes with forecast sensitivity curves and published upper limits at $f_{x} = 10$ with the sensitivity thresholds and upper limits of the four radio telescopes.}
    \label{fig:fig3a}
\end{figure}

Fig. \ref{fig:fig3a} indicates the detectability of 21SSD models at $f_x = 10$ against the sensitivity and the upper limits of the four radio telescopes considered. HERA and GMRT are forecast to be sensitive to all models along the entire range of wavenumbers offered. MWA possesses more restricted behavior; it does not detect any of the models for $k < 0.06 \ h\,\mathrm{Mpc}^{-1}$, but beyond this, it is able to fully detect all models. LOFAR is much more limited. None of the models are visible to LOFAR at $k > 0.16 \ h\,\mathrm{Mpc}^{-1}$. Its most extreme detection window is for models $f_\alpha = 2$ and $r_{H/S} = 0$ or $0.5$, which are still visible to $k = 0.16$. There is a bit less to the model $r_{H/S} = 1$ and $f_\alpha = 2$, which is only visible to $k = 0.09 \ h\,\mathrm{Mpc}^{-1}$. The most restricted scenario is observed in models with $f_\alpha = 1$ and $r_{H/S} = 0$ or $0.5$, with the smallest range of detection among all the models under testing.

\begin{figure}
    \centering
    \includegraphics[width=\linewidth]{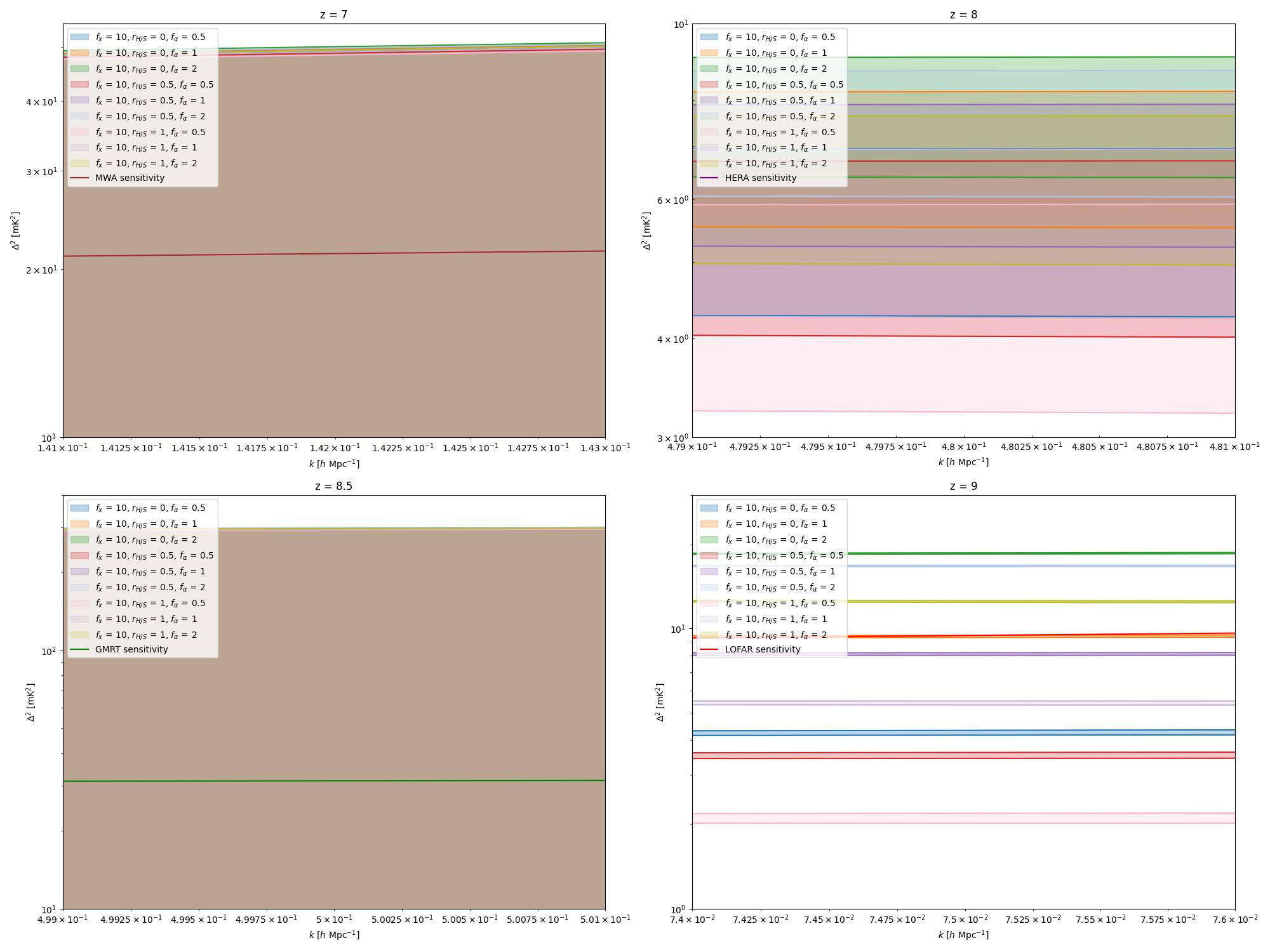}
    \caption{Thermal-noise bands indicate the expected uncertainty at the representative wavenumber of the 21SSD models at $f_x = 10$.}
    \label{fig:fig3b}
\end{figure}

Fig. \ref{fig:fig3b} shows the influence of thermal noise on the distinguishability between the 21SSD models for $f_x = 10$. As the left panels illustrate, both the GMRT and the MWA are noise-limited for model discrimination at the selected wavenumber since their thermal noise is much larger compared to the predicted amplitudes of signal. The same restriction is found for HERA; even though the upper right panel illustrates the model prediction well separated, model discrimination is limited by thermal noise at $k = 0.48 \ h\,\mathrm{Mpc}^{-1}$ and hence no discrimination between models is possible. On the other hand, LOFAR exhibits good distinguishability between models. As the lower right panel illustrates, for $k = 0.075 \ h\,\mathrm{Mpc}^{-1}$, the error bands do not overlap at the plotted representative wavenumber.


\section{Conclusion}
In recent years, radio observations provide essential constraints on the high-redshift 21-cm signal, due to their impact on increasing our understanding of higher redshift observations. Since there is still no fully constrained model for the thermal history of the IGM, whether it is a cosmological model or an astrophysical model, observations of higher redshifts are invaluable. A lot of work has been done in the field of simulating IGM, especially in EoR, such as different types of N-body simulations like Cosmic Reionization On Computers (CROC) \citep{Gnedin_2014}, hydrodynamical simulations like 21SSD, and semi-numerical ones like 21cmFAST, but a noticeable issue is that all these simulations examine different models and none of them can say which model is preferable to other models, only the astrophysical coefficients or various cosmological parameters change, and the obtainable results can be of the type of brightness temperature, density, and so on.

One of the things that all kinds of simulations try to focus on or declare as a result, is the power spectrum, which can be measured in different redshifts. Among the hydrodynamical simulations, 21SSD is used here because of its relatively high spatial and mass resolution. The simulation covers a cube and includes particles, half for the gas and half for dark matter.

Three key parameters are utilized through this simulation, which are the Lyman band emissivity, X-ray emissivity, and X-ray hard-to-soft ratio. In addition to brightness temperature, it obtains the power spectrum, which can be found in the results for each of these models and coefficients. The obtained power spectra from the 21SSD simulations can be compared with the ability of current radio observatories. The sensitivity and thermal noise can be estimated for each of them which has been noticed in recent years.

The detectability of the 21SSD models depends strongly on the adopted instrument, redshift, wavenumber range, and noise assumptions and its instrumental sensitivity. The study underlines the need to perform an effective analysis of the outcome from each telescope, as their ability to distinguish and recognize models presents appreciable variation.  By focusing on the 21SSD simulations, this study presents a method of associating observational results with simulation factors so as to determine models compatible with existing data as well as potential data.

A non-detection would not by itself rule out the simulated physical framework: either the simulations systematically misestimate the amplitude of the 21-cm power spectrum, or the true astrophysical conditions fall outside the parameter space explored by the 21SSD models. In the latter case, alternative simulations such as LoReLi hydrodynamical simulation \citep{meriot2024loreli} could be investigated in future work to broaden the comparison. This simulation allows accurate simulations of the EoR at $256^3$ resolution and it consists of 300 cMpc boxes. Compared to the 21SSD simulation, which performs at $1024^3$ resolution and in a 200 Mpc.$h^{-1}$ box, LoReLi has a lower resolution but a larger simulation box. Since this work is mainly about the power spectrum analysis, higher resolutions are strongly preferred.

In contrast, if detections are made within the simulated ranges, it will become possible to address a deep astrophysical question: the level of X-ray production during the epoch of reionization. By comparing observed 21-cm signals and the results presented in this paper, it will be possible to constrain the X-ray emissivity and thus gain a better insight into the thermal history of the intergalactic medium, evolution of the neutral hydrogen fraction, and brightness temperature fluctuations.

Overall, the six figures from the previous section, which make a direct comparison between the output of the 21SSD simulations and the capabilities of current radio telescopes (MWA, HERA, GMRT and LOFAR in this work), indicate that the questions that we want to answer are not purely theoretical. Instead, they relate to what we can in fact observe with today's instruments, and what we cannot, if the thermal history of the universe is indeed one of the scenarios explored in these simulations.
For the $f_x = 0.1$ model, where the 21-cm signal is not yet very strong, all four telescopes are in principle capable of detecting the signal. However, HERA and GMRT are above the sensitivity curve for almost all models and for almost the entire range of wavenumbers, and can therefore be considered close to ideal. MWA is only successful in an intermediate wavenumber window, while LOFAR can only truly detect the simulated signal in a narrow band around wavenumbers of order $k \sim 0.1\ h\,\mathrm{Mpc}^{-1}$ up to approximately $k \sim 0.142\ h\,\mathrm{Mpc}^{-1}$.

As $f_x $ rises to 1, HERA and GMRT continue to probe nearly the full wavenumber range in the simulations, but the feasible detection window for MWA shrinks and LOFAR’s window constricts significantly. For $f_x = 10$, the comparison between the simulations and the sensitivity curves still shows that HERA and GMRT can detect all models over the whole relevant wavenumber range. In contrast MWA is sensitive only to wavenumbers larger than about $k \sim 0.06\ h\,\mathrm{Mpc}^{-1}$ and LOFAR is effectively limited to a narrow window between about $k \sim 0.09\ h\,\mathrm{Mpc}^{-1}$ and $k \sim 0.16\ h\,\mathrm{Mpc}^{-1}$.

In the second part of the comparison, in particular in Figures 2, 4 and 6, the real thermal noise of these telescopes is displayed more explicitly. This allows a more realistic assessment of how distinguishable the simulated models are in practice. HERA at $k \sim 0.48\ h\,\mathrm{Mpc}^{-1}$ and LOFAR at $k \sim 0.1\ h\,\mathrm{Mpc}^{-1}$ not only detect the simulated signal above their sensitivity limits for $f_x = 0.1$ and $f_x = 1$, but the separation between model curves is large enough compared to the noise that statistically significant discrimination between different 21SSD scenarios would also be possible in real data. By contrast, GMRT and MWA are noise-limited at the representative wavenumbers shown, such that differences between models are effectively unobservable. These instruments are only able to constrain if a non-zero power spectrum is detected or not.
For $f_x = 10$, contrary to what might be expected from the simulation–sensitivity comparison alone, we find that although HERA and GMRT still detect all models in terms of raw sensitivity, the thermal noise at the shown wavenumbers becomes so large that the power of model selection is effectively lost. In contrast, LOFAR at $k \sim 0.075\ h\,\mathrm{Mpc}^{-1}$ is the only telescope in this simulation–detection comparison that keeps the error bands of all models separated, allowing all 21SSD scenarios with $f_x = 10$ to be distinguished from one another within a small but exceptionally strong wavenumber window.
Therefore, when the 21SSD simulations are considered alongside the sensitivity curves and noise properties of current instruments, they indicate that at the redshifts and wavenumbers considered here, HERA and GMRT provide the broadest forecast sensitivity coverage, MWA plays a scale-dependent role, and LOFAR - despite its limited detection window - may be valuable for model discrimination in restricted wavenumber intervals at specific wavenumbers, particularly at high values of $f_x$.


\begin{acknowledgments}
      We express our sincere gratitude to Seyed Sajad Tabasi from Sharif University of Technology in Tehran, Iran, for sharing the foundational idea that inspired this paper, contributing the concept behind the code algorithm, and for guiding us toward the research direction that made this work possible. This work is partially supported by the Iran National Science Foundation (INSF).
\end{acknowledgments}

\nocite{*}
\bibliography{aipsamp}

\end{document}